\documentclass[aps,prx,twocolumn,longbibliography,superscriptaddress]{revtex4-2}

\usepackage{amsmath}
\usepackage{amssymb}
\usepackage{graphicx}
\usepackage[colorlinks,allcolors= blue]{hyperref}
\usepackage{siunitx}
\usepackage{booktabs} 

\renewcommand{\vec}{\boldsymbol}

\begin{document}
\title{Squeezing light with optomechanical and spin-light quantum interfaces}

\author{Gian-Luca Schmid}
\affiliation{Department of Physics and Swiss Nanoscience Institute, University of Basel, 4056 Basel, Switzerland}
\author{Manel Bosch Aguilera}
\affiliation{Department of Physics and Swiss Nanoscience Institute, University of Basel, 4056 Basel, Switzerland}
\author{Chun Tat Ngai}
\affiliation{Department of Physics and Swiss Nanoscience Institute, University of Basel, 4056 Basel, Switzerland}
\author{Maryse Ernzer}
\affiliation{Department of Physics and Swiss Nanoscience Institute, University of Basel, 4056 Basel, Switzerland}
\author{Luiz Couto Correa Pinto Filho}
\affiliation{Center for Macroscopic Quantum States, bigQ, Department of Physics, Technical University of Denmark, 2800 Kongens Lyngby, Denmark}
\affiliation{Danish Fundamental Metrology, 2970 Hørsholm, Denmark}
\author{Dennis Høj}
\affiliation{Center for Macroscopic Quantum States, bigQ, Department of Physics, Technical University of Denmark, 2800 Kongens Lyngby, Denmark}
\author{Ulrik Lund Andersen}
\affiliation{Center for Macroscopic Quantum States, bigQ, Department of Physics, Technical University of Denmark, 2800 Kongens Lyngby, Denmark}
\author{Florian Goschin}
\affiliation{Institut für Experimentalphysik, Universität Innsbruck, Technikerstraße 25, 6020 Innsbruck, Austria}
\author{Philipp Treutlein}
\affiliation{Department of Physics and Swiss Nanoscience Institute, University of Basel, 4056 Basel, Switzerland}

\begin{abstract}
We investigate squeezing of light through quantum-noise-limited interactions with two different material systems: an ultracold atomic spin ensemble and a micromechanical membrane. Both systems feature a light-matter quantum interface that we exploit, respectively, to generate polarization squeezing of light through Faraday interaction with the collective atomic spin precession, and ponderomotive quadrature squeezing of light through radiation pressure interaction with the membrane vibrations in an optical cavity. Both experiments are described in a common theoretical framework, highlighting the conceptual similarities between them. The observation of squeezing certifies light-matter coupling with large quantum cooperativity, a prerequisite for applications in quantum science and technology.
In our experiments, we obtain a maximal cooperativity of $C_\mathrm{qu} =10$ for the spin and $C_\mathrm{qu} = 9$ for the membrane.
In particular, our results pave the way for hybrid quantum systems where spin and mechanical degrees of freedom are coherently coupled via light, enabling new protocols for quantum state transfer and entanglement generation over macroscopic distances.
\end{abstract}

\maketitle

\section{Introduction}
Light-matter quantum interfaces are paradigmatic systems for the study of quantum measurements, quantum feedback control, and quantum networking \cite{hammerer2010quantum}. Such interfaces enable precise measurements of quantum systems \cite{sewell2012magnetic, rossi2019observing, pezze2018quantum}, the creation and storage of non-classical states of light \cite{heshami2016quantum}, and the generation of entanglement between remote and even very different systems \cite{thomas2021entanglement}. 
For many experiments and applications, a quantum-noise-limited light-matter interface is essential.  Achieving this necessitates that the coupling rate between the light and the matter system surpasses the decoherence rate, which ensures that the interaction remains robust against noise and dissipation. 

The condition for a quantum-noise-limited interface is expressed in terms of the \textit{quantum cooperativity}
\cite{aspelmeyer2014cavity, bowen2016optomechanics}
\begin{equation}
    C_\mathrm{qu} = \frac{S_\mathrm{qba}}{S_\mathrm{th}},
\end{equation} 
defined as the ratio between the quantum backaction noise $S_\mathrm{qba}$ imparted by the light on the system due to their interaction and the thermal noise $S_\mathrm{th}$ driving the system due to its coupling to the environment, which is the main decoherence source considered here. A quantum-noise-limited operation thus requires $C_\mathrm{qu}>1$.

In many experiments, the quantum cooperativity of the interface is extracted or calibrated by analyzing the scaling behavior of various noise sources, for example, with the power of the probe light. As optical power increases, so does backaction noise \cite{aspelmeyer2014cavity}, providing direct evidence of quantum backaction and validating the high-cooperativity regime. This method relies on the correct calibration of the different contributions to the measured signal, as well as the assumption that all technical noise sources are well below the backaction level. A more direct approach for certifying large quantum cooperativity is the observation of the quantum correlations induced on the light through its interaction with the system. These correlations redistribute noise between orthogonal light quadratures, resulting in squeezing of a specific quadrature of the light. This squeezing is a definitive indicator of strong light-matter coupling, as it directly reveals the quantum nature of the interaction.

The observation of optical squeezing as a means to quantify the quantum cooperativity is broadly applicable across various light-matter systems. For mechanical oscillators, this phenomenon is commonly referred to as \textit{ponderomotive squeezing} \cite{mancini1994quantum, fabre1994quantum}. Ponderomotive squeezing was observed for silicon nitride membranes at cryogenic temperatures \cite{purdy2013strong} and more recently even at room temperature \cite{huang2023room}, for waveguide resonators \cite{safaviNaeini2013squeezed}, single levitated nano-particles \cite{militaru2022ponderomotive, magrini2022squeezed}, and atomic ensembles in a Fabry-Pérot cavity \cite{brooks2012nonclassical}. More recently, this technique has been applied to other types of quantum light-matter interfaces. For example, the polarization state of the light was squeezed through its interaction with the collective spin of an atomic ensemble, both at acoustic frequencies \cite{jia2023acoustic} and at megahertz frequencies \cite{baerentsen2024squeezed}. These developments underscore the usefulness of squeezing as a tool for probing and characterizing light-matter interactions in both microscopic and macroscopic systems.

In this manuscript we will introduce the squeezing mechanism in general terms and suggest that it can be used to characterize a broad range of light-matter interfaces. To exemplify this, we present experimental demonstrations of squeezing of light through interactions with two very different systems: the collective spin precession of an ensemble of ultracold atoms and the vibrations of a micromechanical membrane.

In section~\ref{sec:General}, we present the general framework for quantum-noise-limited measurements of a harmonic oscillator and discuss how squeezing of light is generated and how it can be detected. In the following two sections, we demonstrate squeezing of specific light quadratures in microscopic and macroscopic systems using two distinct experimental platforms: In section~\ref{sec:Spin} the framework is applied to the interface between the collective spin of a cold atomic cloud and a detuned probe laser beam; in section~\ref{sec:Membrane} a micromechanical oscillator in an optical cavity is used to squeeze the light. The observation of squeezing in both systems confirms their quantum-noise-limited operation. In section \ref{sec:Conclusion} we point out that this is a prerequisite for coupling the systems in a quantum coherent fashion using light as a mediator of interactions.

\begin{figure*}[t]
\includegraphics{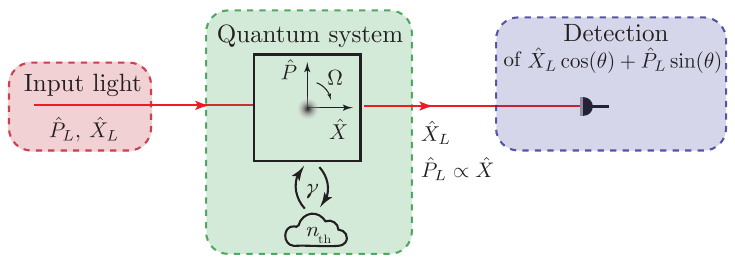}
\caption{Schematic of the light-matter quantum interfaces discussed in this paper: Coherent light interacts with a quantum system and is detected after the interaction. The quantum system is modeled as a harmonic oscillator of frequency $\Omega$, which is damped and stochastically driven through its coupling at rate $\gamma$ to a thermal environment. The light-matter interaction maps the $\hat X$ quadrature of the oscillator onto the $\hat P_L$-quadrature of the output light, while the $\hat X_L$-quadrature of the input light drives the oscillator, representing the measurement backaction. By adjusting the homodyne angle $\theta$, any superposition of the output light quadratures can be detected. }
\label{fig:system_ligh_interface}
\end{figure*}

\section{Quantum-noise-limited measurements}
\label{sec:General}

We begin by providing a general description of our light-matter interfaces, where light interacts with systems modeled as harmonic oscillators characterized by an angular frequency $\Omega$ and an energy decay rate $\gamma$, see Fig.~\ref{fig:system_ligh_interface}. These systems are also coupled to a thermal environment at temperature $T$, with an average occupation number $n_\mathrm{th} = [\exp(\hbar \Omega/k_\mathrm{B}T)-1]^{-1}$. This coupling results in a total decoherence rate given by $\gamma_\mathrm{th} = \gamma(n_\mathrm{th} + 1/2)$.

The interaction between light and the oscillator is described by a Hamiltonian in which a light quadrature $\hat{Q}_{L,i}$ linearly couples to one of the oscillator's quadratures $\hat{Q}_{j}$, where the light quadratures are $\hat{\vec{Q}}_{L} = (\hat{X}_L, \hat{P}_L)$ and the oscillator quadratures $\hat{\vec{Q}} = (\hat{X}, \hat{P})$. The interaction Hamiltonian is of the form
\begin{equation}\label{eq:generalH}
\hat{H}_\mathrm{int} = \hbar g \, \hat{Q}_{L,i} \hat{Q}_{j},
\end{equation}
where the coupling constant $g$ has units of [\unit{s^{-1/2}}] and $\hat{Q}_{L,i}$ is a quadrature of a traveling light field with units of [\unit{s^{-1/2}}] satisfying the commutation relation $[\hat{X}_L(t), \hat{P}_L(t')] = \mathrm{i} \delta(t-t')$. The system's operators are dimensionless and satisfy the commutation relation $[\hat{X}, \hat{P}] = \mathrm{i}$.

 As an example, consider the specific coupling $\hat{H}_\mathrm{int} = \hbar g \hat{X}_{L} \hat{X}$. In this case, the Langevin equations of motion for the system quadrature operators are given by \cite{clerk2010introduction}
    \begin{align}
    \label{eq:ofMotionX}
         & \partial_t{\hat{X}}(t) = \Omega \hat{P}(t), \\
         & \partial_t{\hat{P}}(t) = -\Omega \hat{X}(t) - \gamma \hat{P}(t) +\sqrt{2\gamma}\hat{P}_\mathrm{th}(t) - g\hat{X}_L(t).
    \label{eq:ofMotionP}
    \end{align}
In the absence of coupling to the light, the system is driven solely by a random force $\hat{F}_\mathrm{th} = \sqrt{2\gamma}\hat{P}_\mathrm{th}$, which arises from coupling to the thermal environment. This thermal noise is fully characterized by its power spectral density (PSD) $S_\mathrm{th}(\omega)$, which we assume to be flat around the system's resonance frequency, i.e., $S_\mathrm{th}(\omega \simeq \Omega) = \gamma_\mathrm{th}$. When the coupling to the light is introduced, quantum noise (shot noise) of the light field also drives the system with a force $\hat{F}_\mathrm{qba} = -g\hat{X}_L$. This force represents the unavoidable quantum backaction exerted by a probe during a measurement. The quantum backaction is also characterised by its PSD $S_\mathrm{qba}(\omega)$, which we also assume to be flat around resonance, i.e., $S_\mathrm{qba}(\omega) \simeq S_\mathrm{qba}(\omega \simeq \Omega) = g^2/4$. As  customary, we  define the \textit{measurement rate} $\Gamma = g^2 / 4$ \cite{bowen2016optomechanics, karg2020light, ernzer2023opticalThesis}, and in this case $S_\mathrm{qba}= \Gamma$.

The steady-state dynamics of the system can be characterized by its PSD. From the Langevin equations given in \eqref{eq:ofMotionX} and \eqref{eq:ofMotionP}, we obtain the following PSD (see appendix \ref{appendix:PSD} for a detailed derivation)
\begin{equation}
S_{XX}(\omega) = 2|\chi(\omega)|^2\left[S_\mathrm{th} + S_\mathrm{qba}\right] = 2|\chi(\omega)|^2\gamma_\mathrm{th}(1 + C_\mathrm{qu}),
\label{eq:SXX}
\end{equation}
where $\chi(\omega) = \Omega/(\Omega^2 - \omega^2 - \mathrm{i}\gamma\omega)$ is the system's susceptibility \cite{bowen2016optomechanics}. Here, we have introduced the quantum cooperativity $C_\mathrm{qu} = S_\mathrm{qba}/S_\mathrm{th} = g^2/(4\gamma_\mathrm{th}) = \Gamma/\gamma_\mathrm{th}$ \cite{verhagen2012quantum}. If the quantum cooperativity exceeds unity, the quantum noise of the light (the probe) drives the system more strongly than the thermal noise, placing the system in the so-called quantum-noise-limited regime.

Continuing with the example, $\hat{H}_\mathrm{int}$ maps the $\hat{X}$ quadrature of the system onto the $\hat{P}_L$ quadrature of the output light, which can be used to observe the dynamics of the system. Consequently, if we denote the $\hat{P}_L$ quadrature operator before the interaction as $\hat{P}^\mathrm{(in)}_L$, then after the interaction, we can express it using the input-output relations \cite{gardiner2000quantum, clerk2010introduction} as
\begin{equation}
\hat{P}^\mathrm{(out)}_L = \hat{P}^\mathrm{(in)}_L + \sqrt{4\Gamma}\hat{X}.
\end{equation}
Here, one can see that $\Gamma$ is the rate at which information about the system's quadrature $\hat{X}$ is imprinted onto the output phase quadrature of the light $\hat{P}^\mathrm{(out)}_L$. A measurement of $\hat{P}^\mathrm{(out)}_L$ then yields the power spectral density
\begin{equation}
S^{(\mathrm{out})}_{P_L,P_L}(\omega) = S^{(\mathrm{in})}_{P_L,P_L} + 4\Gamma S_{XX}(\omega),
\label{eq:SPP}
\end{equation}
where the first term $S^{(\mathrm{in})}_{P_L,P_L} = 1/2$ is the shot noise of the light, and the second term is the signal from the system imprinted onto the light. 

For small measurement rates $\Gamma$, the spectrum of the output light is dominated by the first term (shot noise). This shows that a sufficiently large measurement rate $\Gamma$ is required to ensure  that the measurement is not overwhelmed by shot noise, allowing observation of the system's dynamics. However, increasing the coupling strength also enhances the quantum backaction in $S_{XX}$, causing the system to be driven more strongly by the light. Indeed, using Eq.\,\eqref{eq:SXX} we can write 
    \begin{equation}
       S_{P_L,P_L}^\mathrm{(out)}(\omega) =  \frac{1}{2} + 8|\chi(\omega)|^2\gamma_\mathrm{th}\left(\Gamma+\frac{\Gamma^2}{\gamma_\mathrm{th}}\right).
       \label{eq:SPPthreeterms}
    \end{equation}
Here we see that, in a first stage, when the measurement rate is increased, the thermal noise of the system becomes the dominant contribution to the spectrum, which scales linearly with the measurement rate $\Gamma$. However, in the regime of large cooperativity, the dominant noise contribution is the backaction noise, scaling quadratically with the measurement rate $\Gamma$. By carefully calibrating how the noise in the measured spectrum scales for different measurement rates, these differences can be used to estimate the quantum cooperativity.   

A closer look at the light-matter interaction reveals that the interaction generates correlations between the quadratures of the outgoing light field, a phenomenon known as squeezing.  When light interacts with the system, the amplitude quadrature $\hat{X}_L$ of the light drives the system, whose $\hat{X}$ quadrature is in turn imprinted onto the phase quadrature $\hat{P}_L$ of the light. This mechanism leads to correlations between the orthogonal quadratures of the output light field, leading to a redistribution of noise between them. Specifically, the noise in a particular linear combination of the output light quadratures is reduced (squeezed), while the noise in the orthogonal linear combination is increased (anti-squeezed), consistent with the Heisenberg uncertainty principle.

The squeezing of the light can be observed by homodyne detection, which allows to measure any linear combination of quadratures
\begin{equation}
\hat{D}_\mathrm{\theta} = \hat{X}_L \cos\theta + \hat{P}_L \sin\theta,
\end{equation}
where $\theta$ is the \textit{homodyne angle}. 
By adjusting $\theta$, it is possible to measure either the squeezed quadrature (minimum noise) or the anti-squeezed quadrature (maximum noise). We also note that because the system is a harmonic oscillator, squeezing will be frequency-dependent, shaped by the susceptibility of the oscillator $\chi(\omega)$. Near the resonance frequency $\Omega$, the correlations between the light's quadratures are strongest, enabling precision measurements below the shot-noise limit. Far from resonance, the orthogonal quadratures are uncorrelated and the noise of the light field is given by shot noise. Achieving significant squeezing requires a large measurement rate $\Gamma$, which simultaneously enhances quantum backaction and drives the system more strongly with the light. The PSD of an arbitrary output light quadrature $D_\theta$ at an angle $\theta$ is derived in appendix \ref{app:PonderomotiveSqueezing} and given by
\begin{align}\label{eq:generalSDD}
 S_{D_\theta D_\theta}(\omega) = \frac{1}{2} +4\Gamma \Big\{\mathrm{Re} & [\chi(\omega)]\cos(\theta)\sin(\theta)\nonumber \\
  &+  S_{XX}(\omega)\sin^2(\theta)\Big\},
\end{align}
where, as before, the first term is the shot noise of the light, the second term describes the interference between the shot noise on the $\hat X_L$ quadrature and the signal of the driven system on the $\hat P_L$ quadrature, and the third term is the signal of the driven system on the $\hat P_L$ quadrature. The squeezing of the light can therefore be observed by measuring an intermediate quadrature $\theta \neq 0, \pi$ and appears as a dip in the spectrum below the shot-noise level of $1/2$.

In the following, these concepts are investigated with two physically very different but conceptually similar systems: We experimentally demonstrate  squeezing of light by exploiting first its interaction with the spin of a cold atomic cloud (a microscopic system), and then with a micromechanical membrane (a much more macroscopic system).

\section{Spin-light interface}
\label{sec:Spin}

\begin{figure*}[t]
\includegraphics{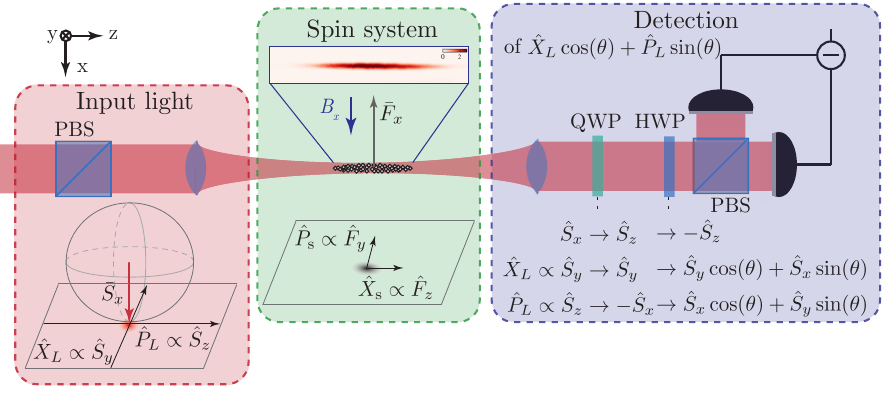}
\caption{Setup of the spin-light interface: The coherent input light is well polarised along $\bar S_x$. In this limit, the polarisation state of the light can be described in terms of harmonic oscillator quadratures $\hat X_L$ and $\hat P_L$. The light is focused on a pencil-shaped cloud of Rubidium atoms and interacts with the atomic spin via the Faraday interaction. Subsequently, the light is detected by polarisation homodyne detection using two waveplates and a polarizing beam-splitter (PBS). The quarter-wave plate (QWP) rotates the local oscillator onto the circular polarisation $\bar S_x \rightarrow \bar S_z$. The half-wave plate (HWP) is then used to set the homodyne angle $\theta$.}
\label{fig:spin_ligh_interface}
\end{figure*}

\subsection{Description of the spin system}
Our spin system is a dense cloud of $N_a = 2.0 \,(2)\times10^7$ cold rubidium atoms trapped in a far-detuned optical dipole trap, with a large optical depth $d_0\simeq 500\,(50)$ along the long axis of the cloud, see Fig.\,\ref{fig:spin_ligh_interface}. To describe the spin of the entire atomic ensemble, we define a collective spin operator \cite{pezze2018quantum} that is the sum of all individual atomic spins, $\vec{\hat F} = \sum_i \vec{\hat f}^{(i)}$, which is the main degree of freedom of the atomic system. The spin of the atoms is pumped into the hyperfine ground state $|f=2,m_f=-2\rangle$ along a bias magnetic field with a spin polarization of $|\bar F_x| /(2N_a)\geq 0.92$. The pencil-shaped atomic cloud (shown in Fig.\,\ref{fig:spin_ligh_interface}) is interfaced with a mode-matched probe laser \cite{mueller2005diffraction, baragiola2014three, schmid2025hybrid}, red-detuned by $-2\pi\times 30\,\mathrm{GHz}$ from the ${}^{87}\mathrm{Rb}$ $D_2$-line. In these conditions, the collective spin interacts with the light via the Faraday interaction \cite{hammerer2010quantum},
\begin{equation}
\hat H_{s} = \hbar \alpha_1 \hat F_z \hat S_z,
\end{equation}
where $\alpha_1$ is the unitless vector component of the atomic polarisability and $\hat S_z$ is the circularly polarised Stokes vector component of the probe light (where each Stokes vector component describes the difference in photon flux of two orthogonal polarizations with units of [\unit{s^{-1}}]). 

\begin{figure*}[t]
\includegraphics[width = \textwidth]{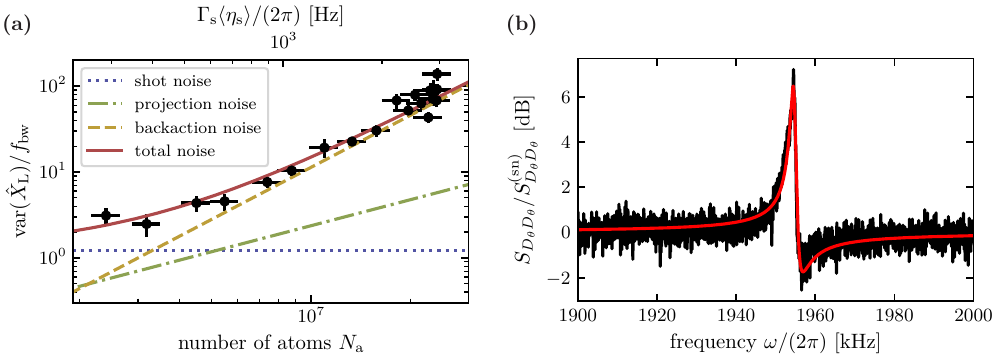}
\caption{(a) Variance of the polarisation fluctuations of the light after the interaction with the atoms. Here, the integration bandwidth is $f_\mathrm{bw} = \Delta_\mathrm{bw}/(2\pi) = \SI{4}{\kilo\hertz}$, which is about an order of magnitude larger than the linewidth of the spin of $\gamma_{s} = 2\pi\times \SI{280}{\hertz}$. The measurement rate is varied by changing the number of atoms in the dipole trap. Each data point is an average over ten measurements. The theory curve is calculated without free parameters, taking the inhomogeneous spin-light coupling into account, as described in the text, see Eq.\,\eqref{eq:spinVar}. Here, the probe light is \SI{-30}{\giga\hertz} red detuned and has a power of \SI{1}{\milli\watt}. The spin oscillator has a frequency of $\Omega_s = 2\pi\times\SI{0.98}{\mega\hertz}$, set by a magnetic bias field of $B_x = \SI{1.4}{G}$. The detection efficiency of $\eta_\mathrm{det} = 0.83$ is included in the theory curve, which increases the effective shot noise contribution from $1$ to $1/\eta_\mathrm{det}$ (Eq.\,\ref{eq:spinVar} is modified accordingly). (b) Spectrum of the light around the spin resonance of $\Omega_s = 2\pi\times\SI{1.958}{\mega\hertz}$ (magnetic bias field of $B_x = \SI{2.8}{G}$). Here, we detect at a homodyne angle of $\theta = 0.19\pi$. Polarization squeezing below the shot noise level is observed due to the quantum-noise limited interaction of the light with the atomic spin. }
\label{fig:spin_meas}
\end{figure*}

For a well-pumped spin along the x-axis, $|\bar F_x| = |\langle \hat F_x \rangle| \gg \sqrt{\langle \hat F_y^2 \rangle}, \sqrt{\langle \hat F_z^2 \rangle}$, the transverse spin components $\hat F_y$ and $\hat F_z$ can be mapped onto harmonic oscillator quadratures via the Holstein-Primakoff approximation \cite{hammerer2010quantum}, $\hat X_{s} = \hat F_z/\sqrt{|\bar F_x |}$ and $\hat P_{s} = \hat F_y/\sqrt{|\bar F_x |}$. The frequency of this spin oscillator $\Omega_{s}$ is given by the Larmor precession frequency, which is set by the bias magnetic field $B_x$. Here, we choose to implement the spin as a positive frequency oscillator, but alternatively a negative frequency oscillator could be implemented as e.g.\ in \cite{baerentsen2024squeezed}. 

The Holstein-Primakoff approximation can also be applied to the polarization Stokes vector of the light if the coherent probe light beam is well polarized. For input light linearly polarised along $\bar S_x = \langle \hat S_x \rangle$, we define the polarisation amplitude and phase quadratures as $\hat X_L^\mathrm{(pol)} = \hat S_y/\sqrt{\bar S_x}$ and $\hat P_L^\mathrm{(pol)} = \hat S_z/\sqrt{\bar S_x}$, respectively, with dimensions [\unit{s^{-1/2}}]. Under these approximations, the Faraday spin-light interaction can be rewritten in terms of harmonic oscillator quadratures as in Eq.\,\eqref{eq:generalH},
\begin{equation}\label{eq:spinH}
    \hat H_{s} = \hbar \sqrt{4\Gamma_{s}}\hat X_{s} \hat P_L^\mathrm{(pol)},
\end{equation}
with the spin measurement rate given by 
\begin{equation}
 \Gamma_{s} = \frac{\alpha_1^2 |\bar S_x||\bar F_x|}{4}.
 \label{eq:measurementrate}
\end{equation}
We find that $\Gamma_s$ depends on the length of the mean spin, the intensity of the probe light and the atomic polarisability constant. 

In order to correctly calibrate the spin-light interface, we have to consider that the radial waist of the 3D-Gaussian atomic cloud of $w_a = \SI{25}{\micro\meter}$ is similar in size to the waist of the Gaussian probe beam $w_0 = \SI{50}{\micro\meter}$. This causes the spin-light coupling to be inhomogeneous. To account for this, the local intensity of the light at the position of the atoms has to be averaged over the ensemble. The normalized mean intensity of the light seen by the atoms is  $\langle \eta_{s} \rangle = \sum_i |u_0(\vec{r}_i)|^2/N_a$ and the normalized mean squared intensity reads  $\langle \eta_{s}^2 \rangle = \sum_i |u_0(\vec{r}_i)|^4/N_a$, which has to be taken into account for the calibration of spin noise properties \cite{mueller2005diffraction,hu2015entangled}. Here, $\vec{r}_i$ is the position of the $i$th atom and the laser mode function $u_0(\vec{r})$ is normalised to unity at the focus $u_0(\vec{0})=1$. The effective spin quadratures are then defined as $\hat X_{s,\mathrm{eff}} = \frac{\langle \eta_{s} \rangle}{\sqrt{\langle\eta_{s}^2\rangle}}\hat X_{s}$ and $\hat P_{s,\mathrm{eff}} = \frac{\langle \eta_{s} \rangle}{\sqrt{\langle\eta_{s}^2\rangle}}\hat P_{s}$, and the measurement rate as $\Gamma_{s,\mathrm{eff}} = \langle\eta_{s}^2\rangle\Gamma_s$. For our experimental implementation, we estimate $\langle \eta_{s} \rangle = 0.53$ and $\langle \eta_{s}^2 \rangle = 0.33$, which is described in more detail in \cite{schmid2025hybrid}.

\subsection{Scaling with the atom number}

In a first set of experiments, the scaling of the different contributions to the fluctations of the light after the interaction with the spin is studied. For this, the light beam containing the spin signal is measured using polarisation homodyne detection (see Fig.\,\ref{fig:spin_ligh_interface} with $\theta = 0$). For our well-polarized spin, imperfect optical pumping corresponds to a thermal occupation of $\bar{n}_\mathrm{th}=0.03$ \cite{julsgaard2003entanglement} and thus $\gamma_\mathrm{th} \approx \gamma_s/2$, meaning that the spin noise is purely quantum mechanical with $ S_\mathrm{th}(\omega) = \gamma_{s}/2$. Often, this term is called \textit{projection noise} in the literature because it describes the quantum uncertainty of the spin state that appears when it is projected by a measurement. The backaction of the light acting on the spin is as before given by the spin measurement rate $ S_\mathrm{qba}(\omega) = \Gamma_{s,\mathrm{eff}}$. Integrating the recorded spectrum in Eq.\,\eqref{eq:SPPthreeterms} over the spin resonance (with $\Delta_\mathrm{bw}\gg \gamma_{s}$), we obtain the variance of the measured light quadrature \cite{clerk2010introduction}
\begin{align}
\label{eq:spinVar}
    \mathrm{var}(\hat X_L^\mathrm{(pol)}) &= 2\int_{\Omega_{s} - \Delta_\mathrm{bw}/2}^{\Omega_{s} + \Delta_\mathrm{bw}/2}  S_{X_LX_L}(\omega)\,\frac{\mathrm{d}\omega}{2\pi} \nonumber\\
    &= \frac{\Delta_\mathrm{bw}}{2\pi} + 2\Gamma_{s,\mathrm{eff}}\left(1+\frac{2\Gamma_{s,\mathrm{eff}}}{\gamma_{s}}\right),
\end{align}
where the first term is the shot noise of the light, the second term is the projection noise (or thermal noise in case $\bar n_\mathrm{th}\neq 0$) of the spin, and the third term is the backaction noise. Here, we consider $S_{X_LX_L}$ instead of $S_{P_LP_L}$ because of the different quadratures involved in the coupling Hamiltonian given in Eq.\,\eqref{eq:spinH} compared to the example given in section\,\ref{sec:General}. 

In the measurement shown in Fig.\,\ref{fig:spin_meas}\,(a), the measurement rate is varied by loading the dipole trap with a different number of atoms $N_a$ for each data point. This changes the mean spin length $|\bar F_x | \simeq 2N_a$ and therefore the measurement rate $\Gamma_{s,\mathrm{eff}}$ [see Eq.\,\eqref{eq:measurementrate}]. The output of the homodyne detector is measured with a lock-in amplifier. The measurement rate of each individual measurement [the horizontal axis of Fig.\,\ref{fig:spin_meas}\,(a)] is calibrated by aligning the mean spin with the propagation axis of the light and measuring the dc Faraday rotation of the light. 

The theory reproduces the measured variances well. However, the theory lines depend strongly on the correct estimation of the cloud geometry and the linewidth of the spin oscillator. A less calibration-dependent experiment can be performed by using the spin system to squeeze the light, as described in the following section.

Note that the cooperativity of the spin system $C_{s} = \Gamma_{s,\mathrm{\mathrm{eff}}}/\gamma_\mathrm{th} \simeq 2\Gamma_{s,\mathrm{\mathrm{eff}}}/\gamma_{s}$ can be directly read off from the ratio of the backaction noise and the thermal noise. From the comparison of the measurement with the theory we deduce that the backaction noise is up to an order of magnitude larger than the thermal noise, which gives a spin cooperativity of order $C_{s} \approx 10$ for the largest measurement rates in Fig.\,\ref{fig:spin_meas}\,(a).

\begin{figure*}[t]
\includegraphics{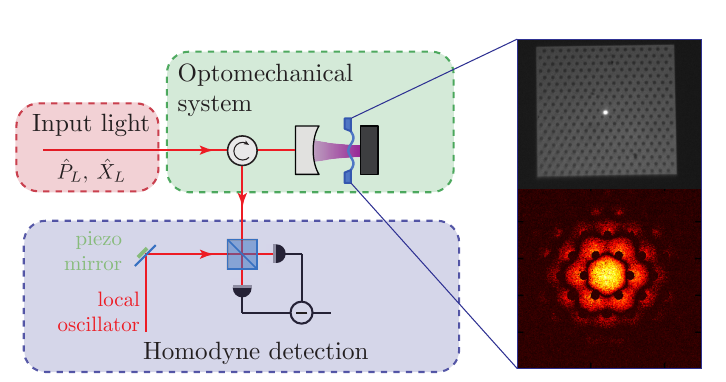}
\caption{The optomechanical interface: Coherent input light drives an optical cavity with a nanomechanical membrane inside, interacting with the membrane vibrations through radiation pressure. An image of the membrane (on top) and the vibration mode of the central defect (below) are shown as insets on the right. The light leaving the cavity is measured by homodyne detection. For this, it is combined with a local oscillator which is derived from the driving laser. The phase between the driving beam and the local oscillator can be controlled with a mirror glued on a piezo crystal. By changing this phase, any superposition between the amplitude quadrature $\hat X_L$ and the phase quadrature $\hat P_L$ of the output light can be detected.}
\label{fig:membrane_ligh_interface}
\end{figure*}

\subsection{Polarization squeezing experiment}
In another experiment, we measure the correlations between different polarisation quadratures of the light induced by the interaction with the spin. They arise as the atomic spin is driven by the quantum noise of the circularly polarised component of the light $\hat P_L^\mathrm{(pol)} \propto \hat S_z$,  while the induced spin fluctuations are mapped onto the linear polarisation component $\hat X_L^\mathrm{(pol)} \propto \hat S_y$, assuming the input light is well-polarised along $\bar S_x$. Thus, the interaction with the spin correlates the $\hat S_z$ and the $\hat S_y$ polarization components of the light. In order to measure a linear combination of the two quadratures, two waveplates are installed: A quarter-wave plate (QWP) rotates the mean polarisation of the light to the circular quadrature, $\bar S_x \rightarrow \bar S_z$, $\hat S_y \rightarrow \hat S_y$, and $\hat S_z \rightarrow -\hat S_x$ (see Fig.\,\ref{fig:spin_ligh_interface}). The half-wave plate (HWP) then sets the angle $\theta$ of the detected polarisation component $D_\theta$ \cite{thomas2021calibration}. 

In Fig.\,\ref{fig:spin_meas}\,(b) a measurement of the polarization fluctuation PSD of the light is shown near the spin resonance, demonstrating polarization squeezing of the light by $-1.74\,(5)\SI{}{\decibel}$ below shot noise. The data is fit with a theory curve according to Eq.\,\eqref{eq:generalSDD}, with the linewidth and measurement rate as free parameters. The theory curve also includes the effect of losses in the detection chain (detection efficiency of $\eta_\mathrm{det} = 0.83$), which reduces the observed squeezing. The fit yields a measurement rate of $\Gamma_{s,\mathrm{eff}} = 2\pi\times 812\,(24)\SI{}{\hertz}$. The squeezing measurement was performed in an earlier stage of the experiment\cite{karg2020strong} with only $N_a = 1.0\,(1)\times 10^7$ atoms and a probe geometry slightly different from that of the experiments presented in the previous section, which explains the difference in measurement rate. The fitted linewidth is $\gamma_{s} = 2\pi\times 1.41\,(2)\SI{}{\kilo\hertz}$, which is inhomogeneously broadened by the presence of multiple Zeeman levels in the $F=2$ hyperfine spin manifold. Since this broadening does not add additional noise, it does not change the area of the integrated spectrum shown in Fig.\,\ref{fig:spin_meas}\,(a), but increases the linewidth obtained from the fit in Fig.\,\ref{fig:spin_meas}\,(b). The resulting spin cooperativity in the squeezing measurement is $C_{s} = 1.14 \,(5)$.

\section{Membrane cavity optomechanical interface}
\label{sec:Membrane}

\begin{figure*}[t]
\includegraphics[width =\textwidth]{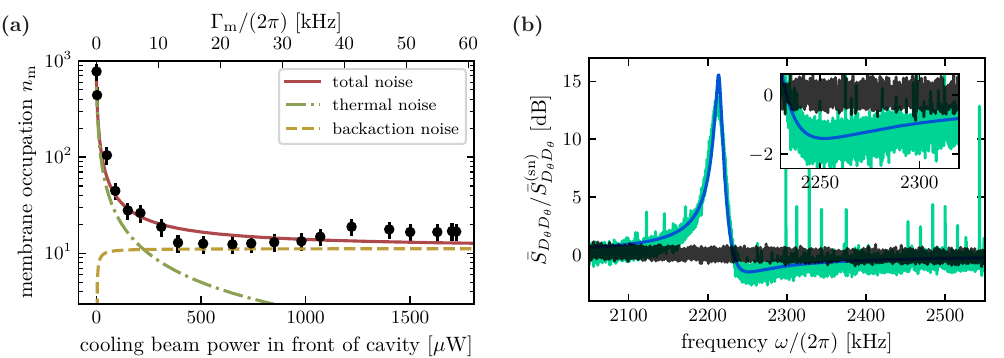}
\caption{(a) Cavity dynamical backaction cooling experiment. The membrane phonon occupation is measured for a red-detuned light beam driving the cavity with different powers. Increasing the optical power decreases the phonon occupation due to thermal noise (dash-dotted line) but increases the contribution from backaction noise (dashed line). For large optical driving powers, the cooling and backaction driving  effects balance and the membrane occupation approaches the theoretical limit of $n_{m} = 11$ phonons. The quantum cooperativity is unity for an input power of $224\,(20)\SI{}{\micro\watt}$ and is about $C_m = 7.6\,(7)$ for the highest applied input power. (b) Spectrum of the light after the interaction with the membrane. The ponderomotive squeezing of the light shows up as a reduction of the noise below the shot noise floor shown in black. The blue line is a fit using Eq.\,\eqref{eq:fullPonderomotiveSqueezing}. The fit yields a linewidth of $\gamma_{m,\mathrm{opt}} = 2\pi\times\SI{5.2}{\kilo\hertz}$ and a measurement rate of $\Gamma_m = 2\pi\times 47\,(2)\SI{}{\kilo\hertz}$, which gives a quantum cooperativity of $C_m = 9.0\,(4)$.}
\label{fig:membrane_meas}
\end{figure*}

\subsection{Description of the membrane system}
In the following we describe experiments where we observe similar physics with an optomechanical system. It consists of a nanomechanical membrane that is patterned with regions of nanopillar arrays forming a phononic crystal \cite{hoej2022ultra}. A defect in the center of the crystal supports a localized vibrational mode with a resonance frequency of $\Omega_{m}=2\pi \times\SI{2.27}{\mega\hertz}$, see Fig.\,\ref{fig:membrane_ligh_interface}. The phononic crystal isolates this vibrational mode from the environment, resulting in a very high mechanical quality factor of $Q_{m} = \Omega_{m}/\gamma_{m}= 5.10\,(3)\times 10^7$ at a cryogenic temperature of $T = \SI{10}{\kelvin}$, determined by a ring-down measurement, see \cite{schmid2025hybrid}. The membrane is embedded in a cavity with linewidth $\kappa = 2\pi\times \SI{94}{\mega\hertz}$. The vibrations of the membrane are coupled to the intracavity field via the optomechanical radiation pressure interaction \cite{aspelmeyer2014cavity}. For our cavity, the two cavity mirrors have different reflectivities ($r_1^2 = 0.995$ and $r_2^2=0.9999$) so that most of the light leaves the cavity through the incoupling mirror. We choose to work deep in the unresolved sideband regime $\kappa\gg\Omega_{m}$ in order to have a fast interaction of the membrane vibrations with the traveling field outside of the cavity. In this regime, the cavity field can be eliminated from the description, allowing the optomechanical interaction to be written as an interaction between the membrane vibrations and the traveling light field outside the cavity,
\begin{equation}\label{eq:membraneH}
    \hat H_\mathrm{om} = \hbar \sqrt{4\Gamma_{m}}\hat X_{m}\hat X_L.
\end{equation} 
Here, $\Gamma_{m} = 4g_\mathrm{om}^2/\kappa$ is the optomechanical measurement rate, with $g_\mathrm{om} = g_0\sqrt{n_c}$ the coherently enhanced optomechanical coupling, $n_c$ the intracavity photon number, and $g_0 = 2\pi\times 248\,(10)\SI{}{\hertz}$ the single-photon single-phonon optomechanical coupling (calibration shown in \cite{schmid2025hybrid}). The unitless membrane position quadrature is given by $\hat X_{m} = \hat x_{m}/(\sqrt{2}x_0)$, with the membrane displacement operator $\hat x_{m}$, the zero-point fluctuation amplitude $x_0 = \sqrt{\hbar/(2m_\mathrm{eff}\Omega_{m})}$, and the effective mass $m_\mathrm{eff}$ of the membrane vibration mode. The Hamiltonian in Eq.\,\eqref{eq:membraneH} is formally equivalent to the Hamiltonian of the atomic system given in Eq.\,\eqref{eq:spinH} and the general Hamiltonian discussed above, given in Eq.\,\eqref{eq:generalH}. 

\begin{table*}
\centering
\begin{tabular}{l|l|l}
\toprule
 & \textbf{Spin-light interface} & \textbf{Optomechanical interface} \\
 \midrule
Coupling mechanism & Faraday interaction & Radiation pressure \\
\midrule
Coupling strength       & $g = \alpha_1 \sqrt{|\bar S_x||\bar F_x|}$       & $g = \frac{4g_0\sqrt{n_c}}{\sqrt{\kappa}}$ \\
 $g = \sqrt{4\Gamma}$    & $\Gamma_s = 2\pi\times 812\, (24)\SI{}{\hertz}$       & $\Gamma_m = 2\pi\times 47\,(2)\SI{}{\kilo\hertz}$ \\
 \midrule
Decoherence rate $\gamma_\mathrm{th}$     & $\gamma_\mathrm{th} \approx \gamma_s/2 = 2\pi\times 705\,(10)\SI{}{\hertz}$       & $\gamma_\mathrm{th} \approx \gamma_m n_m = 2\pi\times 5.21\,(3)\SI{}{\kilo\hertz}$ \\
\midrule
Parameters     &  \begin{tabular}{l@{\hskip 2pt}l}
$\alpha_1$: & Polarisability constant\\
$2|\bar S_x| = \Phi$: & Photon flux\\
$|\bar F_x|=2N_a$: & Collective spin length\\
\end{tabular} & \begin{tabular}{l@{\hskip 2pt}l}
$g_0$: & Single-photon coupling rate\\
$|\bar n_c|$: & Number of cavity photons\\
$\kappa$: & Cavity linewidth\\
\end{tabular} \\
\bottomrule
\end{tabular}
\caption{Main parameters of the two systems in comparison. The coupling strength $g$ [in \unit{s^{-1/2}}] is defined in the Hamiltonian $\hat{H}_\mathrm{int} = \hbar g \, \hat{Q}_{L,i} \hat{Q}_{j}$ given in Eq.\,\eqref{eq:generalH}.}
\label{tab:systems}
\end{table*}

\subsection{Cooling experiment}
First, we consider again the scaling of the different noise contributions that are stochastically driving the mechanical oscillator. For this, we perform a cavity dynamical backaction cooling experiment, in which the optomechanical cavity is driven with a red-detuned laser beam \cite{aspelmeyer2014cavity}. Unlike in the experiments with the atomic spin described above, the light not only drives the membrane vibrations by the quantum backaction of the light, but also reduces its mechanical phonon occupation by providing a viscous damping force. In the unresolved sideband regime, the PSD of the quantum backaction force in an off-resonantly driven cavity is given by
\begin{align}
     S_\mathrm{qba}(\omega) = \frac{g_\mathrm{om}^2}{2}\Big( & \frac{\kappa}{(\kappa/2)^2 + (\Delta_{c}+\omega )^2}\nonumber\\
      &+ \frac{\kappa}{(\kappa/2)^2 + (\Delta_{c}-\omega )^2}\Big),
\end{align}
where $\Delta_{c}$ is the detuning of the driving light from the cavity resonance. While the PSD of the thermal noise of the environment is not affected by the interaction with the light and is simply given by $ S_\mathrm{th} = \gamma_{m}(n_\mathrm{th}+1/2)$, the mechanical susceptibility has to be modified $\chi(\omega)\to\chi_\mathrm{eff}(\omega)$ due to the linewidth broadening and the frequency shift caused by the interaction with the light \cite{aspelmeyer2014cavity}. The frequency is shifted by $\Omega_{m}\rightarrow \Omega_{m} +\delta\Omega_m$ with  $\delta\Omega_m = 2g_\mathrm{om}^2\Delta_{c}/(\kappa^2/4 + \Delta_{c}^2)$, while the membrane linewidth is changed to $\gamma_{m}\rightarrow \gamma_{m} + \gamma_{m,\mathrm{opt}}$ with  $\gamma_{m,\mathrm{opt}} = - 4g_\mathrm{om}^2\Delta_{c}\kappa\Omega_{m}/(\kappa^2/4 + \Delta_{c}^2)^2$ \cite{aspelmeyer2014cavity}. By integrating the mechanical displacement PSD $S_{X_mX_m}$ over the resonance of the membrane, the effective occupation of the membrane oscillator can be calculated, 
\begin{align}
    n_{m} + \frac{1}{2} &= \mathrm{var}(\hat{X}_m) = 2\int_0^\infty S_{X_mX_m}(\omega)\,\frac{\mathrm{d}\omega}{2\pi}\nonumber\\ 
    &= \frac{1}{2} + n_\mathrm{cool} +  \frac{ S_\mathrm{qba}(\Omega_{m})}{\gamma_m + \gamma_{m,\mathrm{opt}}},
\end{align}
where the optically cooled membrane has an effective occupation of $n_\mathrm{cool} = n_\mathrm{th}\gamma_{m}/(\gamma_{m} + \gamma_{m,\mathrm{opt}})$ phonons thanks to the cooling induced by the red-detuned beam, plus a residual heating $n_\mathrm{qba} \approx { S_\mathrm{qba}(\Omega_{m})}/{\gamma_{m,\mathrm{opt}}}$ due to the radiation pressure shot noise of the light.

The mechanical occupation can be measured by performing homodyne detection on the beam reflected from the cavity. The data for such a cooling experiment in a \SI{10}{\kelvin} environment are shown in Fig.\,\ref{fig:membrane_meas}\,(a). We carefully calibrated the homodyne detection chain to convert the detected signal into an occupation number, as described in \cite{ernzer2023optical, schmid2025hybrid}. As the cooling power is increased, the number of phonons in the membrane mode decreases because the membrane is cooled via dynamical backaction. At the same time, the quantum noise on the light drives the membrane, limiting the cooling at high optical powers. In the  unresolved sideband limit this prevents ground-state cooling and limits the phonon occupation to a minimum of $n_{m} = \kappa/(4\Omega_m) \simeq 11$ phonons. We observe that when the cooling beam power is increased, the phonon occupation levels off at $11$ phonons, which is an experimental signature for the operation of the system at high optomechanical quantum cooperativity. In our experiment, the regime of high cooperativity is entered at a driving power of about $224\,(20)\SI{}{\micro\watt}$, above which the drive due to backaction noise exceeds the drive due to thermal noise. For the highest applied laser power, we reach an optomechanical quantum cooperativity of $C_m = 7.6\,(7)$.

\subsection{Ponderomotive squeezing experiment}
The optomechanical system in the quantum regime can also be used to squeeze light. As shown in Eq.\,\eqref{eq:generalSDD}, the squeezing is only observed if the backaction noise is the main driving source of the membrane oscillator. As described above for the general case and for the spin oscillator, in the regime of large cooperativity the driving of the membrane is mainly due to shot noise on the $\hat X_L$-quadrature of the light. This noise is then also mapped onto the $\hat P_L$-quadrature of the light, effectively correlating the amplitude and the phase quadratures. If we perform this experiment, we measure squeezing below shot-noise by $-1.48\,(1) \SI{}{\decibel}$ after the interaction with the membrane, as shown in Fig.\,\ref{fig:membrane_meas}\,(b). To fit the data, we have to take into account that the cavity is driven with a red-detuned beam ($\Delta_{c} = -2\pi\times 40\,(2)\SI{}{\mega\hertz}$, $P = 2.00\,(2)\SI{}{\milli\watt}$), which provides some cooling of the membrane but at the same time mixes the $X_L$ and the $P_L$ quadratures of the light. The model for the fit is therefore more sophisticated than the one in Eq.\,\eqref{eq:generalSDD} and is given in appendix \ref{app:PonderomotiveSqueezing}. The optomechanical quantum cooperativity obtained from this fit is $C_m = 9.0\,(4)$. 

\section{Conclusion and outlook}
\label{sec:Conclusion}

We have reported experiments with a spin-light and an optomechanical quantum interface in the quantum-noise-limited regime. Both systems, although physically very different, can be described in a common framework of a harmonic oscillator whose displacement is coupled to a quadrature of the light field. The main parameters of the two systems are summarised in Tab.\,\ref{tab:systems}. As the coupling strength is increased, we observe that the light-matter interfaces enter the regime of large quantum cooperativity, where the quantum fluctuations of the light are the dominant driving force of the oscillator. Furthermore, the interaction with the oscillator correlates the quantum noise of orthogonal quadratures of the light field, which generates optical squeezing. By observing the squeezing in suitable optical quadratures after the interaction with the atomic spin and the nanomechanical membrane, respectively, we certify the operation of the two interfaces with large quantum cooperativity. These concepts are very general and can be transferred to other systems featuring a light-matter quantum interface, such as various solid-state emitters or atoms in optical cavities.

The two quantum interfaces demonstrated in our experiment open up the possibility to implement various quantum protocols. A particularly interesting perspective is to use the light to mediate a coupling between the atomic spin and the membrane oscillator over a long distance \cite{moller2017quantum,karg2020light,thomas2021entanglement,schmid2022coherent}, exploiting their conceptual similarity even further. For such couplings to operate in the quantum coherent regime, the individual light-matter interfaces have to be quantum noise limited \cite{karg2019remote}.

A light-mediated Hamiltonian coupling between the atomic spin and the mechanical oscillator, which has so far only been realized in a thermal-noise-dominated regime \cite{karg2020light, schmid2022coherent}, can be generated by coupling the two systems with the light in a looped geometry \cite{karg2019remote}, with a phase shift of $\pi$ applied to the quantum signal between the two interactions. In this case, an effective Hamiltonian coupling between the spin and the membrane oscillator of the form  
$
\hat{H}_\mathrm{hyb} = 2\hbar g_\mathrm{hyb} \hat{X}_m \hat{X}_s
$  
can be engineered \cite{karg2019remote}. Here, the coupling rate is determined by the product of the measurement rates of the individual systems to the light,  
$
g_\mathrm{hyb} = ({4\Gamma_m\Gamma_s})^{1/2}.
$
The cooperativity of the hybrid spin-membrane coupling is given by  
\begin{equation}
 C_\mathrm{hyb} = \frac{4g_\mathrm{hyb}^2}{\gamma_{s,\mathrm{tot}} \gamma_{m,\mathrm{tot}}},       
\end{equation}
where the total decoherence rates $\gamma_{i,\mathrm{tot}}$ include both thermal noise driving and decoherence due to backaction noise,  
$
\gamma_{i,\mathrm{tot}} = \gamma_{i,\mathrm{th}} + {\Gamma_{i,\mathrm{ba}}}/{2}.
$
In a looped geometry, backaction noise can be canceled \cite{karg2019remote}, and the hybrid cooperativity is bounded by the product of the cooperativities of the individual systems,  
$
C_\mathrm{hyb} < 16\,C_m C_s.
$
By coupling the two quantum-noise-limited interfaces described in this work, it thus becomes possible to generate quantum coherent interactions between the spin and the membrane. This can be used e.g.\ to entangle the two systems over a distance, or to use the spin ensemble as a coherent controller for the mechanical oscillator \cite{schmid2022coherent}, opening up many exciting opportunities for quantum science and technology.

\section*{Acknowledgement}
G.-L.S., M.B.A., C.T.N, M.E., and P.T. acknowledge funding from the Swiss National Science Foundation (Grant No. 197230), the Swiss Nanoscience Institute (SNI). 
M.B.A. acknowledges funding from the Research Fund for Junior Researchers from the University of Basel.
L.C.C.P.F., D.H. and U.L.A. acknowledge the Danish National Research Foundation (bigQ, DNRF0142) and the Novo Nordisk Foundation (CBQS, NNF24SA0088433).
F.G. acknowledges the Austrian Science Fund (FWF, Grant DOI 10.55776/W1259).

\appendix

\onecolumngrid

\section{From the Langevin equations to the power spectral density}
\label{appendix:PSD}
In this appendix, we present how we calculate the PSD starting from the Langevin equations. The Langevin equations \eqref{eq:ofMotionX} and \eqref{eq:ofMotionP} can be written in the frequency domain as 
\begin{equation}
    \hat X(\omega) = \chi(\omega)\left[\sqrt{2\gamma}\hat P_\mathrm{th}(\omega) - g\hat X_{L}(\omega) \right],
\end{equation}
where the system's susceptibility is given by $\chi = \Omega/(\Omega^2-\omega^2-\mathrm{i}\gamma\omega)$. Here, we have defined the Fourier transform of an operator as
\begin{align}
    \hat O(\omega)=&\frac{1}{\sqrt{2\pi}}\int_{-\infty}^\infty \hat O(t) \mathrm{e}^{\mathrm{i}\omega t}\mathrm{d}t,\\
    \hat O(t)=&\frac{1}{\sqrt{2\pi}}\int_{-\infty}^\infty \hat O(\omega) \mathrm{e}^{-\mathrm{i}\omega t}\mathrm{d}\omega.
\end{align}
In order to calculate the system's PSD, we have to consider the noise properties of the thermal bath and the input light. For here, we assume that we have 
\begin{align}
   &\langle \hat X_\nu (\omega) \hat X_\mu(\omega') \rangle = \langle \hat P_\nu (\omega) \hat P_\mu(\omega') \rangle = \left(n_\nu + \frac{1}{2}\right)\delta(\omega + \omega')\delta_{\nu\mu},\\
    &\langle \hat X_\nu (\omega) \hat P_\mu(\omega') \rangle = -\langle \hat P_\nu (\omega) \hat X_\mu(\omega') \rangle = \frac{\mathrm{i}}{2}\delta(\omega + \omega')\delta_{\nu\mu},
\end{align}
where the indices are given by $\mu, \nu \in \{ \mathrm{L}, \mathrm{th}\}$. For the thermal noise term, $n_\mathrm{th}$ is the thermal occupation of the environment, while for the optical field $n_{L} = 0$. For an operator with stationary statistics, the Wiener–Khinchin  theorem can be applied to obtain the PSD \cite{gardiner2000quantum,clerk2010introduction}
\begin{align}
    S_{XX}(\omega) =& \int_{-\infty}^\infty \langle \hat X (t)\hat X(0)\rangle\,\mathrm{e}^{\mathrm{i}\omega t} \,\mathrm{d}t = \int_{-\infty}^\infty \langle \hat X (\omega)\hat X(\omega')\rangle \,\mathrm{d}\omega'\\
    =&  \int_{-\infty}^\infty \chi(\omega)\chi(\omega') \left[2\gamma \langle \hat P_\mathrm{th} (\omega) \hat P_\mathrm{th}(\omega') \rangle + g^2 \langle \hat X_{L} (\omega) \hat X_{L}(\omega') \rangle \right]\,\mathrm{d}\omega'\\
    =&  |\chi(\omega)|^2\left[ 2\gamma \left(n_\mathrm{th} + \frac{1}{2}\right) + \frac{g^2}{2} \right],
\end{align}
which is the expression given in Eq.\,\eqref{eq:SXX}.

\section{Ponderomotive squeezing of the membrane in a cavity using a red-detuned beam}
\label{app:PonderomotiveSqueezing}
The linearised  Hamiltonian describing the optomechanical interaction  between the membrane and the cavity photons is given by \cite{aspelmeyer2014cavity}
\begin{equation}
\label{mem, eq: Hom linearised}
\hat H_\mathrm{om} = -\hbar g_0\sqrt{n_{c}}\sqrt{2}\hat X_{m}\left(\hat c  + \hat c^\dagger  \right),
\end{equation}
where $n_{c} = \langle \hat c^\dagger \hat c\rangle$ is the average number of photons in the cavity. 
From this Hamiltonian, the following equations of motion can be derived,
\begin{align}
&\partial_t \hat{X}_{m}  = \Omega_{m}  \hat{P}_{m},\\
&\partial_t \hat{P}_{m}  = -\Omega_{m}  \hat{X}_{m} - \gamma_{m} \hat{P}_{m} - \sqrt{2} \left(g_\mathrm{om}^* \hat c +g_\mathrm{om}\hat c^\dagger\right) + \sqrt{2\gamma_{m}} \hat{P}_\mathrm{th},\\
&\partial_t \hat c  = \left(\mathrm{i}\Delta_{c} - \frac{\kappa}{2}\right)  \hat{c}  - \sqrt{\kappa\eta}\hat{a}^\mathrm{(in)} - \mathrm{i}\sqrt{2}g_\mathrm{om}\hat X_{m}.
\end{align}
Here, we have defined a general optomechanical coupling strength $g_\mathrm{om} = g_0 \alpha_L \sqrt{\kappa\eta}/(\kappa/2-\mathrm{i}\Delta)$ which takes the phase difference between the incoming field and the cavity field due to the cavity detuning into account. Furthermore, the cavity incoupling efficiency is given as $\eta = \kappa_1/\kappa$. In the frequency domain, the cavity field is given by
\begin{align}
\hat c(\omega) = -\chi_{c}(\omega)\left(\sqrt{\kappa \eta} \hat{a}^\mathrm{(in)}(\omega) + \mathrm{i}\sqrt{2}g_\mathrm{om}\hat X_{m}(\omega)\right),
\end{align}
where the cavity susceptibility is defined as
\begin{equation}
\chi_{c}(\omega) = \frac{1}{\kappa/2 - \mathrm{i}(\omega+ \Delta_{c})}.
\end{equation}
The ponderomotive squeezing affects the outgoing light, thus we derive the outgoing light quadratures. The outgoing light is given by the incoming field plus the effect of the cavity on the light
\begin{align}
\hat X_{L}^\mathrm{(out)} &= \hat X_{L}^\mathrm{(in)} + \sqrt{\kappa \eta}\hat{X}_{c},\\
\hat P_{L}^\mathrm{(out)} &= \hat P_{L}^\mathrm{(in)} + \sqrt{\kappa \eta}\hat{P}_{c},
\end{align}
with the incoming light quadratures defined as
\begin{align}
 \label{CavityOptoMech eq X_out}
 \hat{X}_{L}^\mathrm{(in)}  &= \frac{1}{\sqrt{2} } ( \hat{a}^\mathrm{(in)\dagger} + \hat{a}^\mathrm{(in)} ), \quad
 \hat{P}_{L}^\mathrm{(in)} = \frac{\mathrm{i}}{\sqrt{2} } ( \hat{a}^\mathrm{(in)\dagger} - \hat{a}^\mathrm{(in)} ),
 \end{align}
and the cavity quadratures as 
 \begin{align}
 \label{CavityOptoMech eq X_c}
 \hat{X}_{c}  &= \frac{1}{\sqrt{2} } ( \hat{c}^\dagger + \hat{c} ), \quad
 \hat{P}_{c} = \frac{\mathrm{i}}{\sqrt{2} } ( \hat{c}^\dagger - \hat{c} ).
 \end{align}
 
Plugging the expression from above in the equation for the outgoing light quadratures, we obtain
\begin{align}
\hat X_L^\mathrm{(out)}(\omega) &= -\kappa \eta \xi_-(\omega)\hat{P}_{L}^\mathrm{(in)} + (1-\kappa\eta \xi_+(\omega))\hat{X}_{L}^\mathrm{(in)} - \eta \kappa \mathcal{R}_-(\omega) g_0\alpha_L\hat{X}_{m}(\omega),\\
\hat P_L^\mathrm{(out)}(\omega) &= \kappa \eta \xi_-(\omega)\hat{X}_{L}^\mathrm{(in)} + (1-\kappa\eta \xi_+(\omega))\hat{P}_{L}^\mathrm{(in)} - \eta \kappa \mathcal{R}_+(\omega) g_0\alpha_L\hat{X}_{m}(\omega),
\end{align}
where we defined
\begin{align}
\xi_+(\omega) &= \frac{\chi_{c}(\omega) + \chi_{c}^*(-\omega)}{2},\\
\xi_-(\omega) &= \mathrm{i}\frac{\chi_{c}(\omega) - \chi_{c}^*(-\omega)}{2},\\
\mathcal{R}_+(\omega) &= \left( \chi_{c}(0)\chi_{c}(\omega) + \chi_{c}^*(0)\chi_{c}^*(-\omega)\right),\\
\mathcal{R}_-(\omega) &= \mathrm{i}\left( \chi_{c}(0)\chi_{c}(\omega) - \chi_{c}^*(0)\chi_{c}^*(-\omega)\right).
\end{align}
The mechanical quadrature can be rewritten in the frequency domain as
\begin{equation}
\hat{X}_{m}(\omega)  = \chi_{m,\mathrm{eff}}(\omega)\left[\eta \kappa \alpha_L g_0 \left(\mathcal{R}_+(\omega)\hat{X}_{L}^\mathrm{(in)} +  \mathcal{R}_-(\omega)\hat{P}_{L}^\mathrm{(in)} \right) + \sqrt{2\gamma_{m}} \hat{P}_\mathrm{th}\right],
\end{equation}
where the effective susceptibility is given by 
\begin{equation}
\chi_{m,\mathrm{eff}}(\omega)^{-1} = \frac{1}{\Omega_{m}}\left(\Omega_{m}^2 - \omega^2 - \mathrm{i}\gamma_{m}\omega - 4|g_\mathrm{om}|^2\Omega_{m}\xi_-(\omega)\right).
\end{equation}
The homodyne detection signal is given by \cite{ernzer2023optical}
\begin{equation}
\hat D(\omega)= \cos(\theta)\hat X_L^\mathrm{(out)}(\omega) + \sin(\theta)\hat P_L^\mathrm{(out)}(\omega), 
\end{equation}
where $\theta$ is the homodyne angle that allows us to adjust the detected light quadrature. Expressing $\hat D(\omega)$ in terms of the input light quadratures and the thermal drive of the membrane, we obtain 
\begin{align}
\label{eq membrane detector signal full}
\hat D(\omega) =  \hat X_L^\mathrm{(in)}&\Big[ \cos(\theta) - \kappa \eta\left(\cos(\theta)\xi_+(\omega) - \sin(\theta)\xi_-(\omega)\right)\nonumber \\
&- (\kappa\eta\alpha_L g_0)^2\chi_{m}(\omega)\left(\cos(\theta)\mathcal{R}_-(\omega)\mathcal{R}_+(\omega) + \sin(\theta)\mathcal{R}_+(\omega)^2\right)\Big] \nonumber\\
+\hat P_L^\mathrm{(in)}&\Big[ \sin(\theta) - \kappa \eta\left(\cos(\theta)\xi_-(\omega) + \sin(\theta)\xi_+(\omega)\right) \\
& - (\kappa\eta\alpha_L g_0)^2\chi_{m}(\omega)\left(\cos(\theta)\mathcal{R}_-(\omega)^2 + \sin(\theta)\mathcal{R}_-(\omega)\mathcal{R}_+(\omega)\right)\Big]\nonumber\\
+\hat{P}_\mathrm{th} &\Big[\eta \kappa \alpha_L g_0\sqrt{2\gamma_{m}}\chi_{m}\left(\cos(\theta)\mathcal{R}_-(\omega) + \sin(\theta)\mathcal{R}_+(\omega)\right)\Big].\nonumber
\end{align}
As it is written here, the output light depends on the input noise of the light and the thermal noise of the membrane. The correlators of the stochastic noise terms are given by
\begin{align}
&\langle \hat X_L^\mathrm{(in)}(\omega) \hat X_L^\mathrm{(in)}(\omega')\rangle = \langle \hat P_L^\mathrm{(in)}(\omega) \hat P_L^\mathrm{(in)}(\omega')\rangle = \frac{1}{2}\delta(\omega + \omega'),\\
&\langle \hat X_L^\mathrm{(in)}(\omega) \hat P_L^\mathrm{(in)}(\omega')\rangle = -\langle \hat P_L^\mathrm{(in)}(\omega) \hat X_L^\mathrm{(in)}(\omega')\rangle = \frac{\mathrm{i}}{2}\delta(\omega + \omega'),\\
&\langle \hat P_\mathrm{th}^\mathrm{(in)}(\omega) \hat P_\mathrm{th}^\mathrm{(in)}(\omega')\rangle = \left(\bar n_\mathrm{th} + \frac{1}{2}\right)\delta(\omega + \omega'),
\end{align}
while the thermal noise and the optical input noise do not correlate. 
We can write this expression as $\hat D(\omega) = A(\omega)\hat X_L^\mathrm{(in)} +  B(\omega)\hat P_L^\mathrm{(in)}+ C(\omega)\hat{P}_\mathrm{th}$. The symmetrised power spectral density of the detected field is then given by 
\begin{equation}
\label{eq membrane PSD full}
 S_{DD}(\omega) = |A(\omega)|^2  S_{XX}^\mathrm{(in)} + |B(\omega)|^2  S_{PP}^\mathrm{(in)} + |C(\omega)|^2  S_{PP}^\mathrm{(th)},
\end{equation}
where the individual noise power spectral densities are given by
\begin{align}
 S_{XX}^\mathrm{(in)} =  S_{PP}^\mathrm{(in)} = \frac{1}{2} \quad \mathrm{and} \quad  S_{PP}^\mathrm{(th)} = n_\mathrm{th} + \frac{1}{2}.
\end{align}
This equation is used to fit the data in Fig.\,\ref{fig:membrane_meas}\,(b).

\paragraph{Resonantly driven cavity: }
Writing all the terms of equation \eqref{eq membrane PSD full} results in a long and not very easily understandable expression. In order to gain a intuitive understanding of the membrane PSD given in equation \eqref{eq membrane PSD full}, we consider the limit of a resonantly driven cavity, i.e. $\Delta_c = 0$. 
In this case, equation \eqref{eq membrane detector signal full} simplifies significantly to
\begin{align}
\hat D(\omega) =  \hat X_L^\mathrm{(in)}&\Big[ \cos(\theta)(1 - \kappa \eta\xi_+(\omega) )- (\kappa\eta\alpha_L g_0)^2\chi_{m}(\omega) \sin(\theta)\mathcal{R}_+(\omega)^2\Big]\nonumber\\
+\hat P_L^\mathrm{(in)}&\Big[ \sin(\theta)(1 - \kappa \eta\xi_+(\omega))\Big]\\
+\hat{P}_\mathrm{th} &\Big[\eta \kappa \alpha_L g_0\sqrt{2\gamma_{m}}\chi_{m}(\omega)\sin(\theta)\mathcal{R}_+(\omega)\Big],\nonumber
\end{align}
which can be re-expressed as
\begin{align}
\hat D(\omega) = &\hat X_L^\mathrm{(in)}\left[\left(1-\chi_{c}(\omega)\kappa\eta\right)\cos(\theta) +4 g_\mathrm{om}^2\kappa\eta\chi_{m}(\omega)\chi_{c}^2(\omega)\sin(\theta)\right]\nonumber\\
&+ \hat P_L^\mathrm{(in)}\left[\left(1-\chi_{c}(\omega)\kappa\eta\right)\sin(\theta)\right]\\
&+\hat{P}_\mathrm{th}\left[2 g_\mathrm{om}\sqrt{\kappa\eta}\chi_{m}(\omega)\chi_{c}(\omega)\sqrt{2\gamma_{m}}\sin(\theta)\right].\nonumber
\end{align}
For a cavity with a very large linewidth $\kappa\gg\Omega_m$ around the membrane resonance $\Omega_m/\omega\simeq 1$, one can write $|\chi_{c}|^2 \rightarrow 4/\kappa^2$. Using this, we can calculate a very simple expression for the symmetrised power spectral density. Here, we apply the Lorentzian approximation and evaluate the power spectral density at $\omega \approx \Omega_{m}$ to get
\begin{align}
 S_{DD} = \frac{1}{2}& - (1-\eta)\eta\left(\frac{\kappa}{2}\right)^2|\chi_{c}(\omega)|^2\nonumber\\
+&4g_\mathrm{om}^2 \eta\kappa|\chi_{m}(\omega)|^2|\chi_{c}(\omega)|^4\nonumber\\
&\cdot\left[\gamma_{m}\kappa(1-\eta)(\Delta_{c}-\omega) + 2\left((2\eta - 1)\left(\frac{\kappa}{2}\right)^2 + (\Delta - \omega)^2\right)(\omega - \Omega_{m})\right]\sin(\theta)\cos(\theta)\\
+&8g_\mathrm{om}^4\kappa^2\eta^2|\chi_{m}(\omega)|^2|\chi_{c}(\omega)|^4 \sin^2(\theta)\nonumber\\
+&8\gamma_{m,\mathrm{th}}g_\mathrm{om}^2\kappa \eta |\chi_{m}(\omega)|^2|\chi_{c}(\omega)|^2\sin^2(\theta).\nonumber
\label{eq:fullPonderomotiveSqueezing}
\end{align}
If we neglect the losses $\eta = 1$, this simplifies to 
\begin{align}
 S_{DD} = \frac{1}{2}& \nonumber\\
+&8g_\mathrm{om}^2 \kappa|\chi_{m}(\omega)|^2|\chi_{c}(\omega)|^2(\omega - \Omega_{m})\sin(\theta)\cos(\theta)\nonumber\\
+&8g_\mathrm{om}^4\kappa^2|\chi_{m}(\omega)|^2|\chi_{c}(\omega)|^4 \sin^2(\theta)\\
+&8\gamma_{m,\mathrm{th}}g_\mathrm{om}^2\kappa |\chi_{m}(\omega)|^2|\chi_{c}(\omega)|^2\sin^2(\theta).\nonumber
\end{align}
If we further assume that the cavity linewidth is large, $\kappa\gg \omega$, we can simplify $\kappa |\chi_{c}(\omega)|\rightarrow 2$ and get
\begin{align}
 S_{DD} = \frac{1}{2}+8\Gamma_{m} |\chi_{m}(\omega)|^2\Big[(\omega - \Omega_{m})\sin(\theta)\cos(\theta) + \left(\Gamma_{m} + \gamma_{m,\mathrm{th}}\right)\sin^2(\theta)\Big].
\end{align}
This is the expression given in Eq.\,\eqref{eq:generalSDD}.

\bibliography{BibLargeCooperativity}

\end{document}